\documentclass[namedreferences]{SolarPhysics}
\usepackage[optionalrh]{spr-sola-addons} % For Solar Physics 
\usepackage{graphicx}        % For eps figures, newer & more powerfull
\usepackage{solaheader}
\usepackage{color}           % For color text: \color command
\usepackage{url}             % For breaking URLs easily trough lines
\newcommand{\aaps}{   {\it Astron. Astrophys. Suppl.}}

\begin{document}

\begin{article}

\begin{opening}

\title{Influence of the Conversion Layer on the Dispersion Relation of Waves in the Solar Atmosphere}

\author{M.~\surname{Haberreiter}$^{1}$\sep
        W.~\surname{Finsterle}$^{2}$
       }
\runningauthor{Haberreiter and Finsterle}
\runningtitle{Dispersion Relation of Waves in the Solar Atmosphere}

   \institute{$^{1}$ Laboratory for Atmospheric and Space Physics, University of Colorado, 1234 Innovation Drive, Boulder, CO, 80303, USA email: \url{haberreiter@lasp.colorado.edu} \\ 
              $^{2}$ Physikalisch-Meteorologisches Observatorium Davos, Dorfstrasse 33, CH-7260 Davos Dorf, Switzerland email: \url{wolfgang@pmodwrc.ch} \\
             }

\begin{abstract}
Observations carried out with the {\it Magneto-Optical Filter at Two Heights} (MOTH) experiment show upward-traveling wave packets in magnetic regions with frequencies below the acoustic cut-off. We demonstrate that the frequency dependence of the observed travel times, {\it i.e.} the dispersion relation shows significant differences in magnetic and non-magnetic regions. More importantly, at and above the layer where the Alfv\'en speed equals the sound speed we do not see the dispersion relation of the slow acoustic mode with a lowered cut-off frequency. Our comparisons with theoretical dispersion relations do not suggest this is the slow acoustic wave type for the upward low-frequency wave. From this we speculate that partial mode conversion from the fast acoustic to the fast magnetic wave might take place. 
\end{abstract}
\end{opening}
%-------------------------------------------------
\section{Introduction}
\label{Introduction} 
Recently, the observation and analysis of upward traveling waves in the solar atmosphere have become of great interest, as they are thought to provide a considerable amount to the heating of the chromosphere. Commonly it was understood that only waves with frequencies above the acoustic cut-off frequency ($\omega_0$) could travel freely in the solar atmosphere, whereas waves with lower frequencies are trapped inside the acoustic cavity of the Sun. However, recent studies by several researchers have detected upward traveling waves with frequencies $\omega < \omega_0$. 

First, from observations with the {\it Transition Region and Coronal Explorer} (TRACE; Handy {\it et~al.}, 1999) De Pontieu, Erd{\'e}lyi, and James (2004) and McIntosh and Jefferies (2006) show that the presence of an inclined solar magnetic field allows low-frequency waves to propagate into the solar chromosphere. The same effect was shown by Jefferies et~al. (2006) with data taken with the {\it Magneto-Optical filter at Two Heights} (MOTH; Finsterle {\it et~al.} 2004a) experiment. Furthermore, Vecchio {\it et~al.} (2007) found low-frequency upward-traveling waves in the velocity signal measured simultaneously in Fe\,{\sc{i}} {7090}\,{\AA} and Ca\,{\sc{ii}} {8542}\,{\AA} with the {\it Interferometric BIdimensional Spectrometer} (IBIS; Cavallini 2006).

What is the nature of the observed low-frequency waves, {{\it i.e.} are these Alfv\'en waves, slow acoustic, or fast magnetic waves}? De Pontieu, Erd{\'e}lyi, and James (2004), McIntosh and Jefferies (2006), and Jefferies {\it et~al.} (2006) consider the low-frequency waves in areas with inclined magnetic field to be the field-guided slow acoustic waves mode (see Section\,\ref{sec:modes} for details) leaking through ``portals'' where the inclined magnetic field lowers the acoustic cut-off. Numerical simulations, carried out by Steiner {\it et~al.} (2007) and Rosenthal {\it et~al.} (2002) indicate the possibility of mode conversion and transmission, which goes in line with theoretical calculations by Cally (2005, 2006, 2007). From MOTH observations, Finsterle {\it et~al.} (2004b) show that the ``magnetic canopy'' ({\it i.e.} the $\beta \approx 1$ layer, with $\beta=8\pi p_{\mathrm{g}}/B^2$, and $p_{\mathrm{g}}$ being the gas pressure and $B$ the {magnetic flux density}) has a strong influence on the propagation behavior of magnetoacoustic-gravity waves. 
\section{Wave Modes}\label{sec:modes}
The different wave modes in the solar atmosphere are very well understood from the theoretical point of view, however the determination of the wave modes from the observation remains still a difficult task. The actual wave type depends on the local sound speed ($c$), Alfv\'en speed ($a$), pressure ($p$), temperature ($T$), magnetic flux density ($B$) and field inclination ($\theta$). Generally, the following types of waves can be distinguished (for details see {\it e.g.} Rosenthal {\it et~al.}, 2002; Bogdan {\it et~al.}, 2002, 2003; Khomenko and Collados, 2006; Cally, 2005, 2006, 2007; Schunker and Cally, 2006). {In high plasma-$\beta$ regions ($a < c$), the fast mode is a longitudinal acoustic wave. 
%In this regime the slow mode is the magnetic wave that is eventually reflected downward. 
In low plasma-$\beta$ ($a > c$) the slow mode is a longitudinal acoustic wave propagating along the magnetic-field lines. Here, the fast mode is the magnetic mode, which, in the case of relatively weak magnetic fields, is nearly aligned with the wave vector irrespective of the field direction. However for high magnetic-field strengths, the fast-mode displacement is perpendicular to $B$ irrespective of the propagation direction (Rosenthal {\it et~al.}, 2002). Finally, the Alfv\'en wave is always transverse to both the magnetic field and the wave vector.}

Stein (1971), Rosenthal {\it et~al.} (2002), Bogdan {\it et~al.} (2002, 2003), and Steiner {\it et~al.} (2007) showed from 2D numerical magneto-hydrodynamics (MHD) simulation that the fast and slow waves undergo coupling at the layer where the plasma-$\beta$ is of the order of unity. Similarly, Cally (2005, 2007) suggested that mode transmission and mode conversion take place at the layer where the sound speed equals the Alfv\'en speed. In Cally's terms, mode transmission is the change of a fast acoustic (magnetic) wave to a slow acoustic (magnetic) wave. Mode conversion is the change from a fast (slow) acoustic to a fast (slow) magnetic wave. In the following we use Cally's notation of mode conversion and transmission. {It is important to note that $\beta=(2/\gamma) c^2/a^2\approx 1.2 c^2/a^2$. This means that the $\beta=1$-layer and the layer where $a=c$ are contiguous. We use the term {\it conversion layer} for the layer where $a=c$. This condition reflects the matching of the phase velocities of the slow and fast waves, which is the fundamental physical precondition for the waves to couple (Cally, 2005).} In order to identify the wave type from the observation ``it is imperative to determine whether the spectral line or continuum is formed in low or high-$\beta$ plasma'' (Bogdan {\it et~al.}, 2003). This is pursued in the present paper with respect to the conversion layer. 
\section{Data}
The MOTH experiment observed simultaneous Doppler shifts in the K\,{\sc{i}} 7699\,{\AA} and Na\,{\sc{i}} 5890\,{\AA} lines during a South Pole campaign in the austral Summer 2002\,--\,03 (Finsterle {\it et~al.}, 2004a) of the full solar disk with a resolution of 3.7\,Mm pixel$^{-1}$. Here we make use of the (455\,Mm $\times$ 455\,Mm 17.8\,hour) data cube taken at disk center starting on 2003 January 20, 00:59 UT. Figure\,1, panel (a) shows the phase travel time $t_\mathrm{ph}$ between the K\,{\sc{i}} 7699\,{\AA} and Na\,{\sc{i}} 5890\,{\AA} observing layers for waves with frequency $\omega$=3.0\,mHz. In non-magnetic regions the mean phase travel time is $\bar{t}_\mathrm{ph}\approx  0$ (green--blue areas), while the red and yellow pixels indicate upward traveling waves in regions that coincide with enhanced {magnetic flux density} in the concurrent MDI magnetogram (panel b). Under the assumption of a significant amount of inclined field lines in the magnetic areas, this agrees well with the findings by De Pontieu, Erd{\'e}lyi and James (2004), as well as McIntosh and Jefferies (2006).
\begin{figure}    
   \centerline{
%\vspace*{-0.1\textwidth}     
\includegraphics[width=0.485\textwidth,height=0.485\textwidth,clip=]{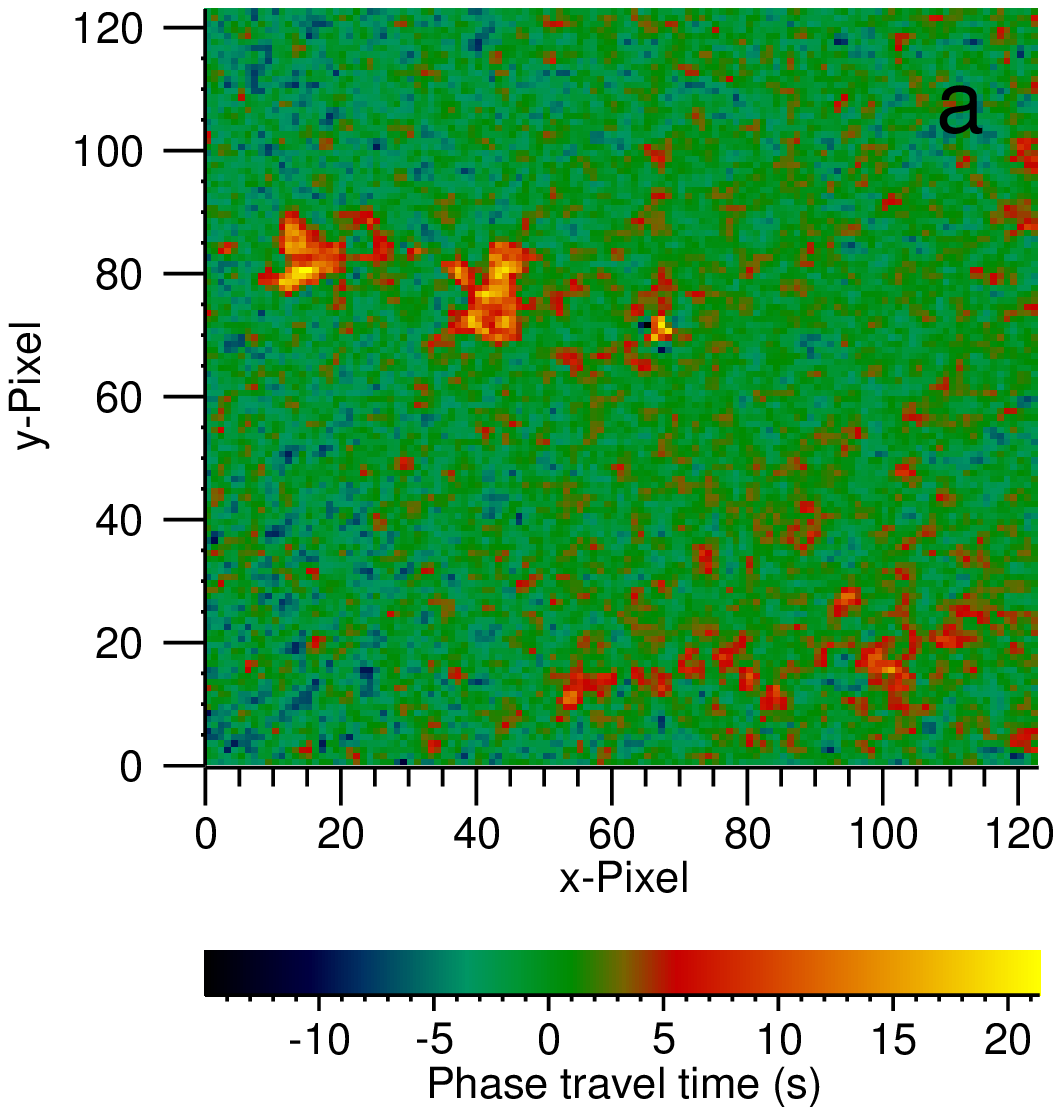}
               \includegraphics[width=0.485\textwidth,height=0.485\textwidth,clip=]{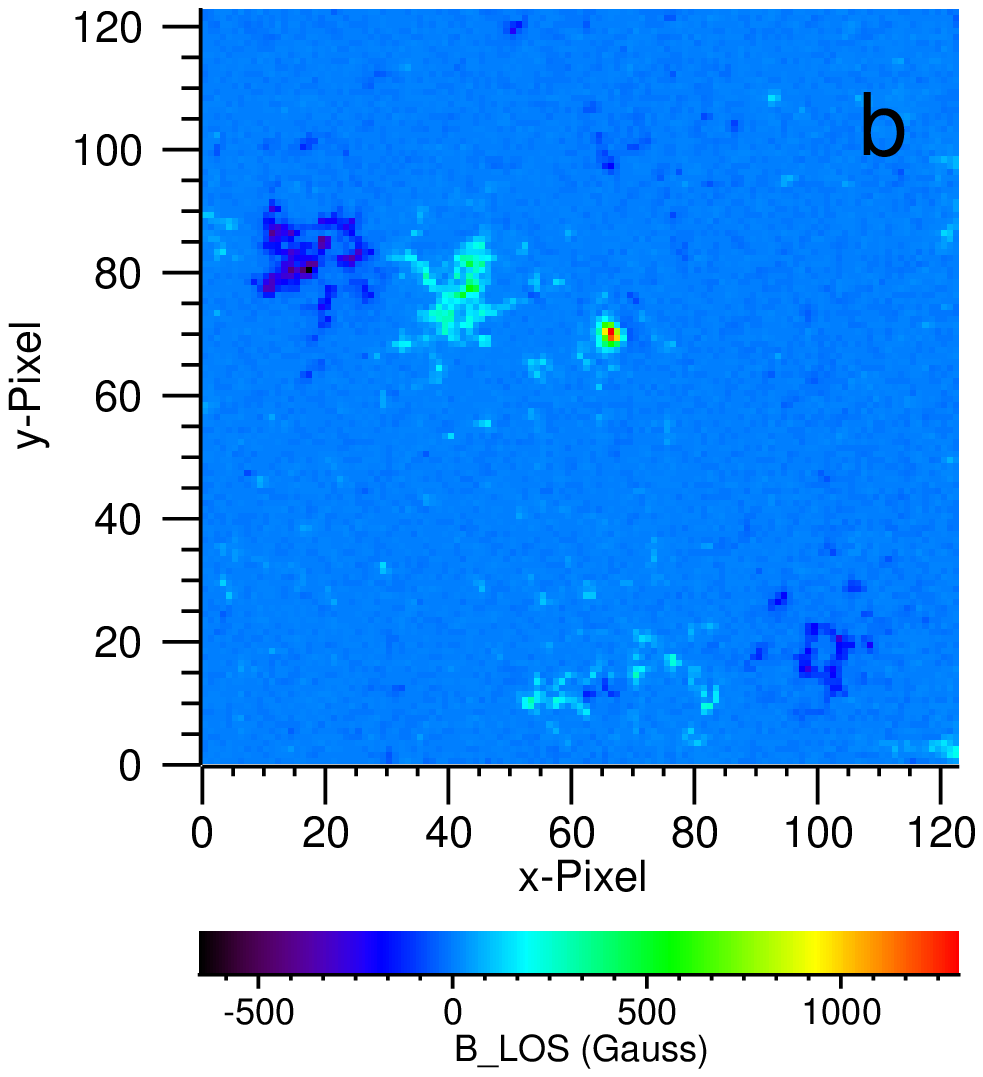}
              }
\caption{Phase travel time map (455\,Mm $\times$ 455\,Mm at disk center) for waves with
frequency $\omega$=3.0\,mHz (a) and the concurrent MDI magnetogram (b). The red and yellow pixels indicate upward traveling waves with phase travel times up to 22 seconds, coinciding with magnetic regions in the MDI magnetogram, whereas $\bar{t}_{\mathrm{ph}}\approx 0$ for the non-magnetic areas.}
\label{fig:ttmap}
\end{figure}
%***********************************************************
\section{Conversion Layer and Formation Heights}
Purely acoustic waves can not propagate below the acoustic cut-off frequency. Following the concept of mode transmission and conversion (Cally, 2007) implies that low-frequency upward-traveling waves only exist in regions where the conversion layer is below the formation heights of the spectral lines employed in the MOTH experiment. Otherwise the low-frequency wave would be evanescent. The reason is that only if the conversion layer is below the height of observation can the upward-propagating wave have coupled to the magnetic field. This goes in line with the concept of {\it portals} by Jefferies {\it et~al.} (2006). To further test this hypothesis we use magnetic-field extrapolations (McIntosh {\it et~al.}, 2001) of the MDI magnetogram in Figure 1b to estimate the plasma-$\beta$ from the plage model atmosphere structure by Fontenla {\it et~al.} (2006). The field extrapolations are carried out with respect to the mean observational height of Ni\,{\sc{i}} {6768}\,{\AA} in the four MDI-filters employed to determine the {magnetic-flux density}. Furthermore, from radiative transfer calculations carried out with COSI (Haberreiter, Schmutz, and Kosovichev, 2008; Haberreiter, Schmutz, and Hubeny, 2008) we determine the formation height of K\,{\sc{i}} 7699\,{\AA} and Na\,{\sc{i}} 5890\,{\AA} at disk center for the quiet Sun and plage. Note that the formation heights change for different view angles or positions on the solar disk. Here we focus only on plage and leave the sunspots for a later study. 

\begin{center}  
\begin{table}[!t]       
    \begin{tabular*}{0.8\textwidth}{cccccc}
      \hline
area        &    $z_{\rm Na}$ (km) & $z_{\rm K}$ (km) & $\Delta z$ (km)& $c$ (km\,s$^{-1}$) & $\omega_{\rm 0}$ (mHz)\\
      \hline
                        \noalign{\smallskip}
%***********************************************************
Quiet Sun&    530  & 160 &  370& 6.5 & 5.6\\
Plage    &    490  & 270 &  220& 7.2 & 5.0\\                       
                        \noalign{\smallskip}
%***********************************************************
      \noalign{\smallskip}
      \hline
  \end{tabular*}
  \caption{Formation height (km) with respect to $\tau_{\mathrm{5000}}=1$ calculated for the quiet Sun and plage for disk center. Also listed are the mean sound speed ($c$) at the formation heights and the cut-off frequency ($\omega_{\rm 0}$) derived from the atmosphere structures.}
\label{tab:formation}
\end{table}
\end{center}
\begin{figure}     \centerline{\includegraphics[width=0.65\textwidth,height=0.55\textwidth,clip=]{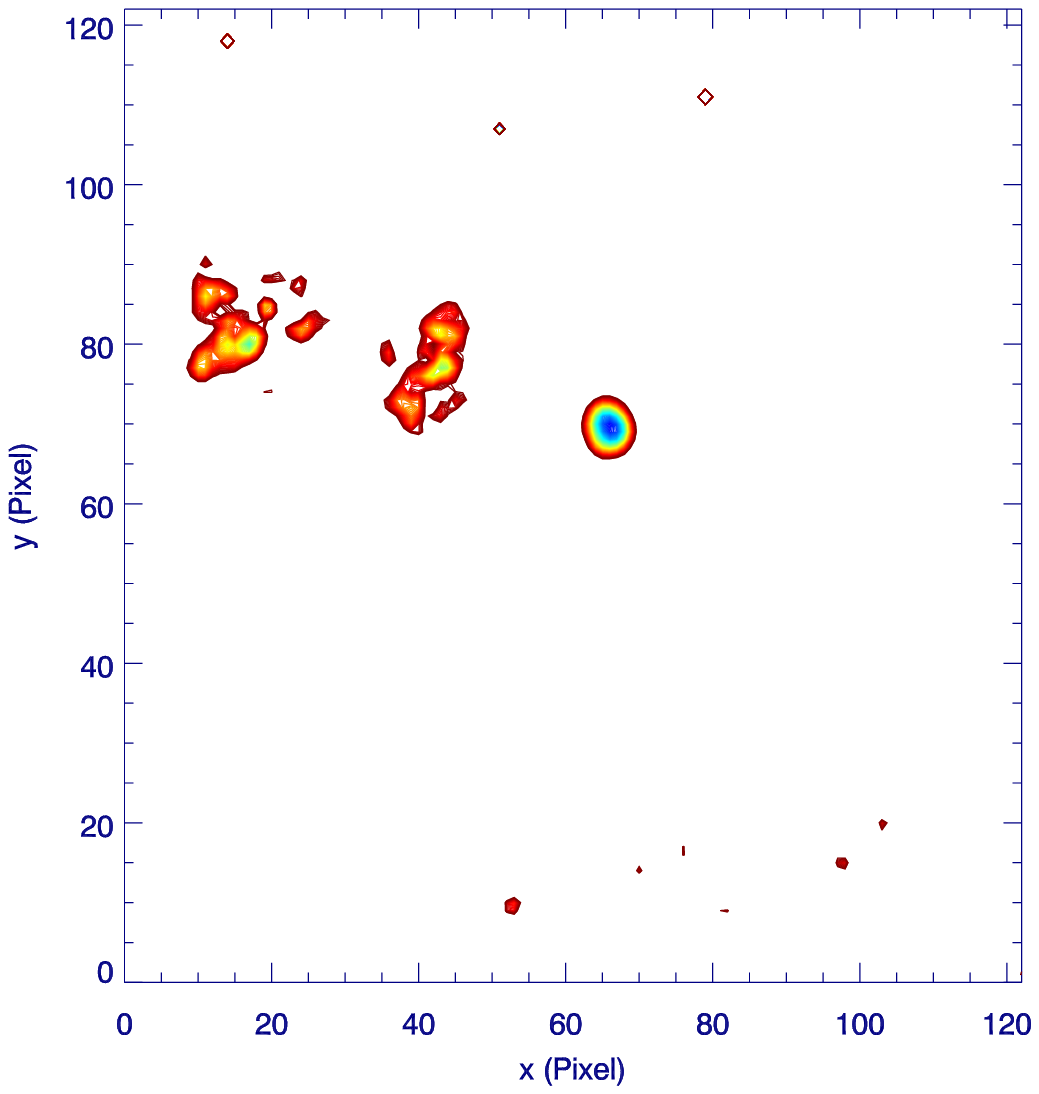}}   \centerline{\hspace*{0.085\textwidth}\includegraphics[width=0.58\textwidth,clip=]{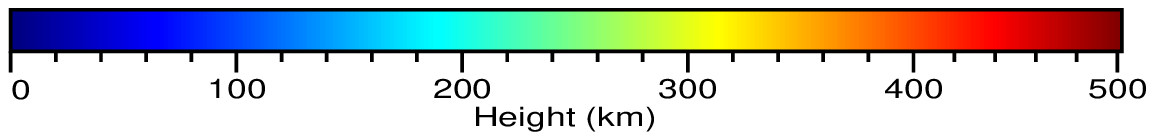}}
\caption{Heights of the conversion layer ($a=c$) where it is below the Na\,{\sc{i}} 5890\,{\AA} formation height in plage with respect to $h(\tau_{\mathrm{5000}}=1)$. The areas where the conversion layer is below the Na\,{\sc{i}} 5890\,{\AA} formation height coincide with areas of higher phase travel time in Figure\,\ref{fig:ttmap}. White indicates that the conversion layer is above the formation height of Na.}
\label{fig:beta}
\end{figure}
From $\tau_{\lambda}=1$ and the position of the Magneto-Optical Filters (MOFs) in the MOTH instrument we derive the mean observing height of K\,{\sc{i}} 7699\,{\AA} and Na\,{\sc{i}} 5890\,{\AA}, given in Table\,\ref{tab:formation}. For a detailed discussion of the MOTH observing heights see Haberreiter, Finstele, and Jefferies (2007). Note that we do not account for any magnetic effects on the formation of the line. This will be pursued in future work. Figure\,\ref{fig:beta} shows the contours of the calculated conversion layer below the Na\,{\sc{i}} 5890\,{\AA} observing height for plage. The red contours indicate where the conversion layer coincides with the observational height of Na\,{\sc{i}} 5890\,{\AA} (490\,km). Toward the center of the active regions the conversion layer drops and reaches heights lower than the observational height of K\,{\sc{i}} 7699\,{\AA}, $z_{\rm K}=$ 270\,km (green to blue contours). The striking congruence of the areas where waves propagate (Figure\,\ref{fig:ttmap}a) and where the conversion layer cuts through the MOTH observing layers, in particular the areas where $z_{a=c}<z_{\rm K}$ supports the assumption by Jefferies {\it et~al.} (2006) that the low-frequency waves propagate only in low-$\beta$ regions. Moreover, this consistency suggests that the formation heights calculated with COSI from the 1D atmosphere structures are correct.
\section{Dispersion Relations}
\begin{figure}    %%%%%%%%%%%%%%%%%% FIGURE 1   
\centerline{\includegraphics[width=.7\textwidth,clip=]{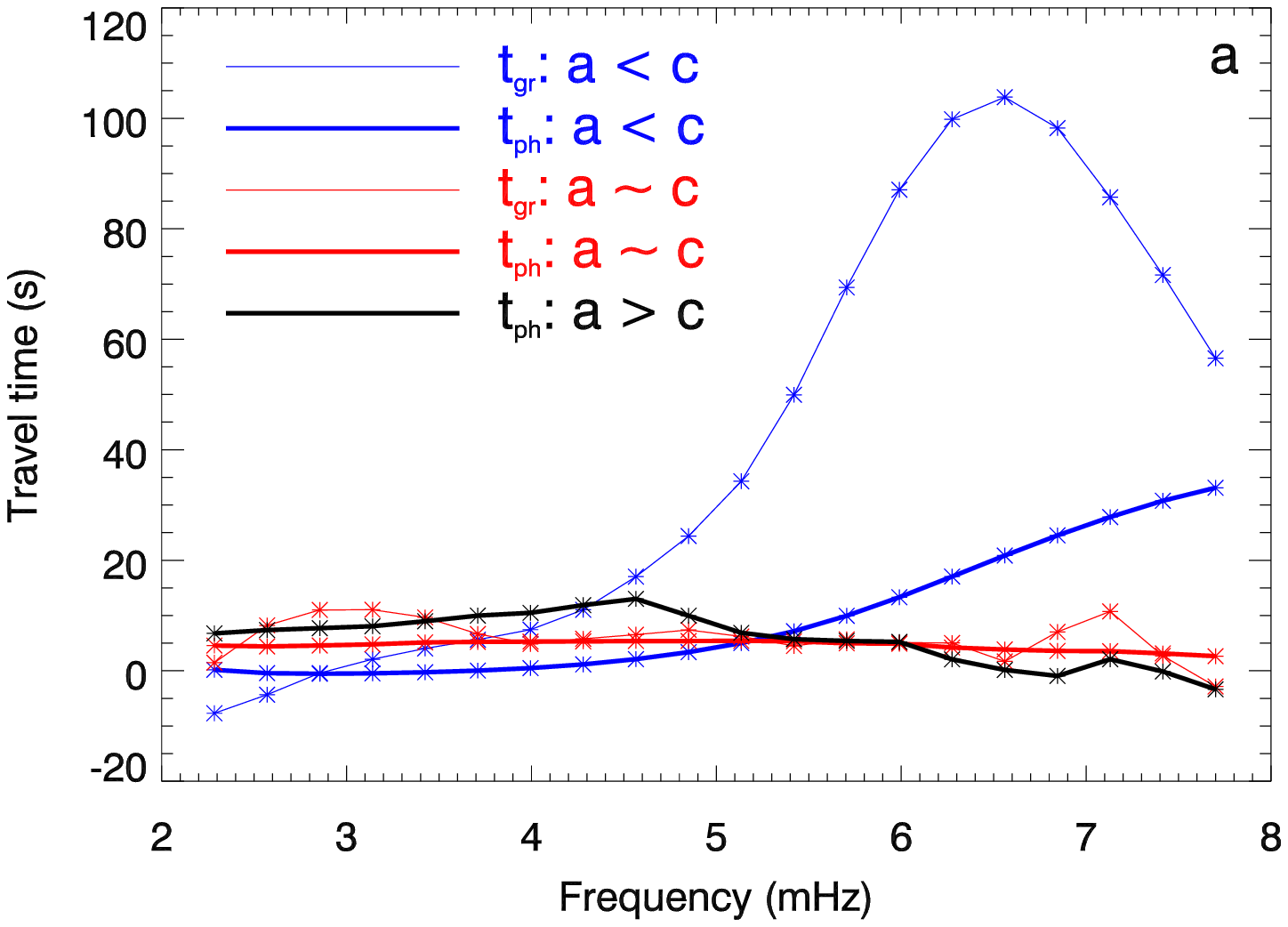}}
\centerline{\includegraphics[width=.7\textwidth,clip=]{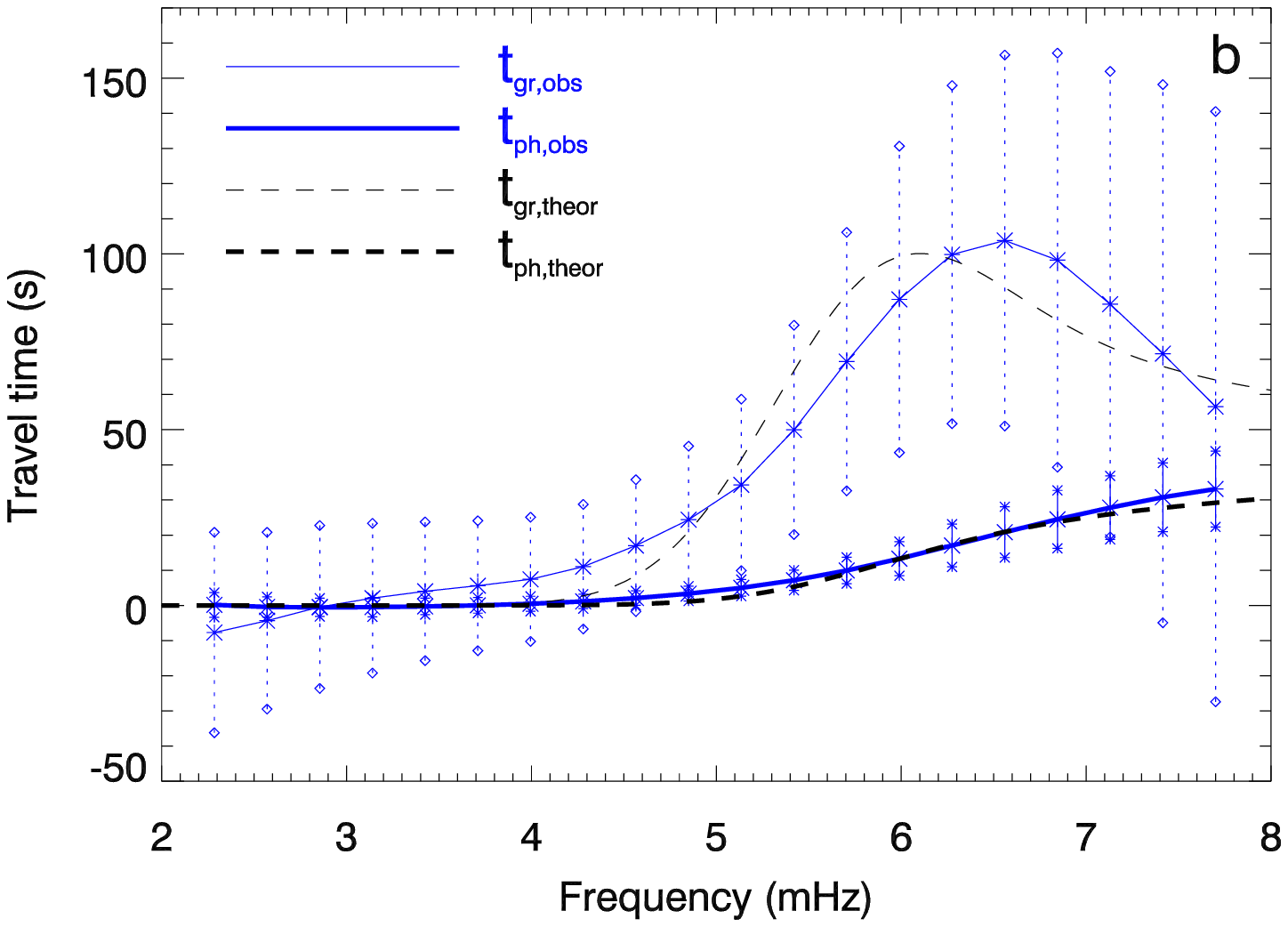}}
\centerline{\includegraphics[width=.7\textwidth,clip=]{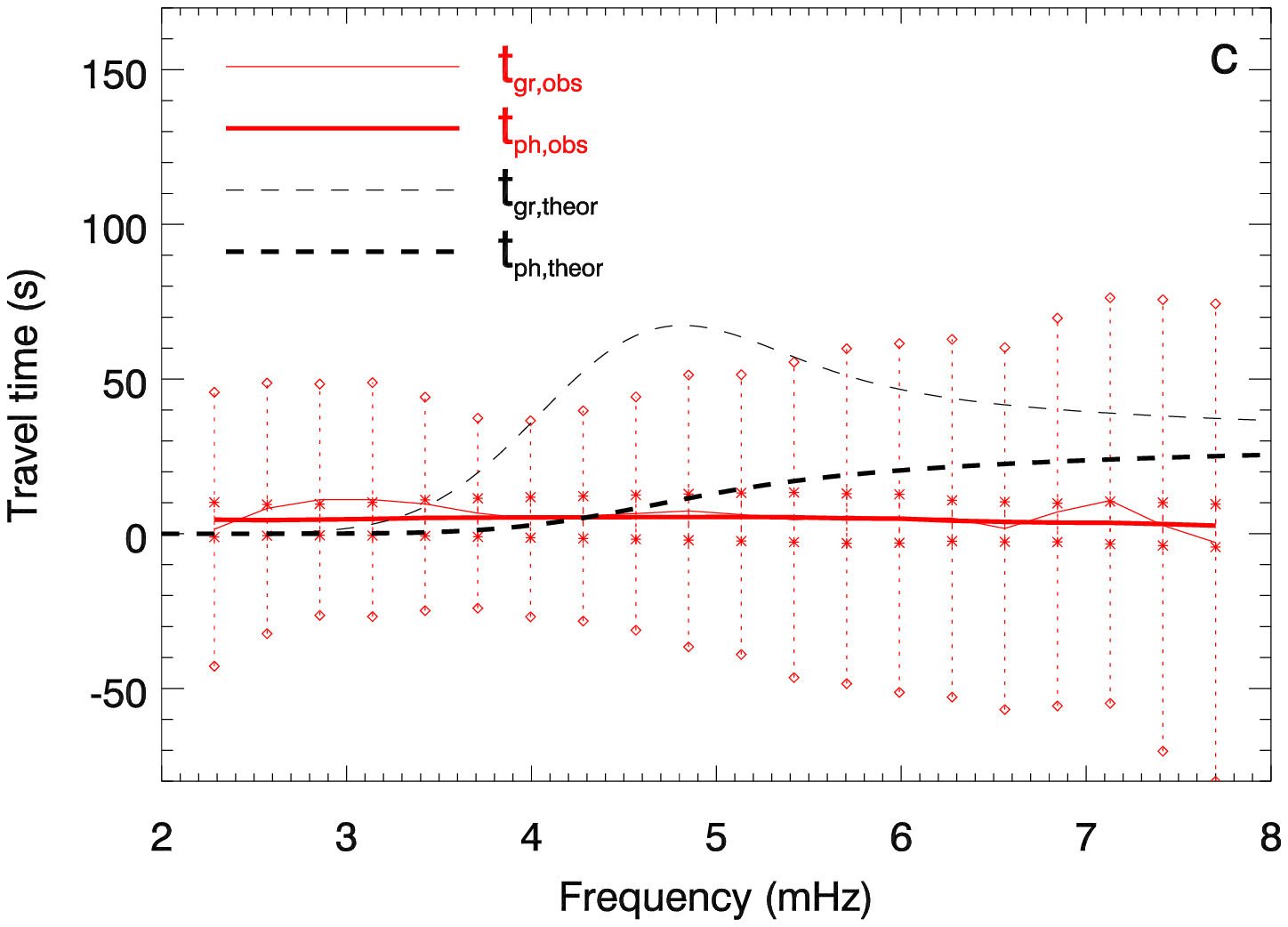}}
\caption{Panel a: Group and phase travel time (thin and thick lines) derived from the MOTH observations as a function of frequency for different height intervals: $a<c$ (blue lines), $a\approx c$ (red lines), and $a>c$ (black line). Panel b: For $a<c$-regime we find the dispersion relation as expected for an acoustic-gravity wave with a cut-off frequency. Also shown are the theoretical travel times (t$_{\mathrm{gr}}$ and t$_{\mathrm{ph}}$) for an acoustic gravity wave with a cut-off frequency $\omega_{\mathrm{0}}=5.6$\,mHz, which agrees well with the observed travel times. Panel c: For the regime where $a\approx c$ we find positive travel times below the acoustic cut-off. Also shown are theoretical dispersion relation based on a lowered cut-off frequency $\omega_{\mathrm{eff}}=\omega_{\mathrm{0}} \cos\theta$, where $\theta=30^{\circ}$ and $\omega_{\mathrm{0}}=5.0$\,mHz. The observed travel times are significantly different from the theoretical values, indicating a different wave type. Panel b and c also show the 1-$\sigma$ error bars for the fitted travel times. Note the different scale of the y-axes.}
\label{fig:disp}
\end{figure}

Low-frequency traveling waves in the solar atmosphere are generally explained by a lower cut-off frequency due to an inclined magnetic field. According to Campos (1987) the reason is that at high altitudes ({\it i.e.} low-$\beta$) the acoustic-gravity waves must follow the magnetic-field lines, increasing the scale height by the factor $(\cos\theta)^{-1}$, with $\theta$ being the field inclination to the vertical. Formally, this can also be described with a change of the effective gravitational acceleration $g_{\mathrm{eff}}=g_{\mathrm{0}}\cos\theta$ (De Pontieu, Erd{\'e}lyi, and James, 2004). A lowering of the acoustic cut-off frequency, however, implies that the propagating wave should still show the acoustic dispersion relation. To test this hypothesis we compare the group and phase travel times derived from the MOTH data with theoretical values. The fitting of phase travel time ($t_{\mathrm{ph}}$) and the group travel time ($t_{\mathrm{gr}}$) is described in detail by Finsterle {\it et~al.} (2004b) and briefly outlined here. In order to study the frequency dependence of the wave propagation, the K and Na data cube is first frequency-filtered in the time domain with a Gaussian of width 0.57~mHz. Note that, as we are interested in the study of the dispersion of propagating wave-packets, the finite width of the frequency filter is necessary to obtain the superposition of waves with different frequencies and at the same time be able to localize the wave packet. The width of 0.57~mHz was chosen, as it allows the localization of a wave-packet with a reasonably well defined frequency. For our analysis we apply a frequency filter between 2.5 and 7.5\,mHz in steps of 0.285\,mHz, indicated by the crosses in Figure\,\ref{fig:disp}a. The frequency-filtered time series are then cross correlated. Finally, the resulting cross-correlation functions are fitted by least-square fits to
\begin{equation}
y(t) = \frac{A^2\delta \omega}{\sqrt{8\pi}}\exp\left[ -\frac{\delta \omega^2 ( t-t_{\mathrm{gr}})^2}{8}\right]  \cos[\omega(t-t_{\mathrm{ph}})]
\end{equation}       
where $t_{\mathrm{ph}}$, $t_{\mathrm{gr}}$, the amplitude ($A$), the central frequency ($\omega$), and the width of the Gaussian filter ($\delta \omega$) are the fitted variables. In this analysis the filter width $\delta\omega$ is allowed to vary as the effective width of the filter is also influenced by the shape of the underlying spectrum. Nevertheless, the fitted widths are typically very close to the given values. The fitted Garbor wavelet function allows us to directly determine the group and phase travel time. We note that a cross-spectrum analysis would be an equivalent procedure to derive these parameters.  

In Figure\,\ref{fig:disp}a the fits for $t_{\mathrm{gr}}$ (thin lines) and $t_{\mathrm{ph}}$ (thick lines) are shown for different height intervals of the $a=c$-layer ($z_{a=c}$) with respect to the formation height of Na ($z_{\rm Na}$) and K ($z_{\rm K}$): \\
{\it i)} $z_{a=c} > z_{\rm{Na}}$ ({\it i.e.} $a < c$, blue lines), \\
{\it ii)} $z_{\rm{Na}} > z_{a=c} > z_{\rm{K}}$ ({\it i.e.} $a\approx c$, red lines), and\\ 
{\it iii)} $z_{\rm{K}} > z_{a=c}$ ({\it i.e.} $a > c$, black line). \\
We find the dispersion relation of acoustic-gravity waves ($\omega_{\mathrm{0}}\approx5.4$\,mHz) for the heights where $a < c$ (blue lines), showing that the criterion $a<c$ is suitable to identify quiet-Sun conditions and that it allows to reproduce the results by Finsterle {\it et~al.} (2004b). {For the regime where $a \approx c$ we find traveling waves at low frequencies ($\omega\le 5$~mHz) with a mean phase travel time of 5.2~seconds (thick red line), and for $a > c$ a mean phase travel time of 9.1~seconds (black line), i.e. with increasing magnetic-flux density the phase travel time increases.} For this analysis we consider areas with magnetic-flux density up to $B=$500~G. The mean magnetic-flux density for the different height intervals are $\bar{B}$=8~G ($z_{a=c} > z_{\rm{Na}}$), $\bar{B}$=109~G ($z_{\rm{Na}} > z_{a=c} > z_{\rm{K}}$), and $\bar{B}$=250~G ($z_{a=c}<z_{\rm{K}}$). Note that as the filling factor for plage regions is smaller than unity, at some areas below the instrumental resolution the ``real'' magnetic flux density is expected to have higher values than $\bar{B}$. The corresponding 1$\sigma$-error bars of the travel time fits are shown in Figure\,\ref{fig:disp}b,c. Moreover, with our approach cannot determine whether the signal observed in one pixel is a superposition of different waves. Due to the fact that the filling factor can be less than one this cannot be ruled out. However, from our analysis it is clear that for the $a~c$ regime the data do to reveal the predominant signal of slow magneto-acoustic gravity waves with a lowered cut-off frequency. 

The group travel time for the $a>c$-regime is not shown due to the increased noise in the fitting of the group travel time combined with poor statistics due to the sparce number of pixels with higher magnetic-flux density. In the following we discribe the comparison of these results with theoretical dispersion relations.

\subsection{Acoustic Wave ($a \ll c$)}
For the fast acoustic (fa) wave we use the dispersion relation 
\begin{equation}
\omega^2=c^2k^2+\omega_0^2, \label{eq:omega} 
\end{equation}
going back to Bel and Leroy (1977), with $c$ being the sound speed, ($k=\alpha + i\beta$) the complex wave number and $\omega_0=c/4H$ is the cut-off frequency for an idealized isothermal atmosphere. Generally, the cut-off frequency for a 1D non-isothermal atmosphere is $\omega_0^2=c^2/(4H^2)(1-2 \mathrm{d} H/\mathrm{d}z)$. As $\mathrm{d}H/\mathrm{d}z$ is small in the layers where Na\,{\sc{i}} 5890\,{\AA} and K\,{\sc{i}} 7699\,{\AA} are formed, we consider the approach by Bel and Leroy (1977) justified. Note, that Equation\,(\ref{eq:omega}) can also be derived from the dispersion relation given by Schunker and Cally (2006), Equation\,(12), for the limit of the non-magnetic case, {\it i.e.} $a \rightarrow 0$. 

{We are aware that the assumption of vertically upward traveling waves {propagating long inclined field lines} is an approximation that is of course not always the case on the Sun. However, for the MOTH instrument the observation of the horizontal component of the wave is limited to small inclination angles. Let us discuss this briefly with the following example. Consider a wave with $\nu=$3.5\,mHz. With a sound speed of 7\,km\,s$^{-1}$ we get a wavelength\,$\lambda=$2000\,km. The horizontal sampling rate ($S$) of the MOTH data is $\approx$3700km\,pixel$^{-1}$. The maximum angle ($\theta_{\rm{max}}$) ($\theta$ being the angle of the propagation vector and the vertical) under which the horizontal component of the wave can still be resolved with MOTH is: $\sin(\theta_{\rm{max}})=\lambda/(2S) \approx 0.27$. From this we get $\theta_{\rm{max}}\approx 16^\circ$. Note that for higher frequencies $\theta_{\rm{max}}$ is smaller, {\it i.e.} for 7\,mHz it is $\approx 8^\circ$. As these angles are small, we consider the 1D approach appropriate.}

From Equation\,(\ref{eq:omega}) we get $k=\pm \sqrt{\omega^2-\omega_0^2}/c$ and the phase velocity:
\begin{equation}
v_{\mathrm{ph,fa}}=\frac{\omega}{\pm k}=\pm \frac{c}{\sqrt{1-\left(\frac{\omega_0}{\omega}\right)^2}} \label{eq:vph}
\end{equation}
Furthermore, using $\omega=\pm \sqrt{c^2k^2+\omega_0^2}$ we get the group velocity:
\begin{equation}
v_{\mathrm{gr,fa}}=\pm \frac{\mathrm{d}\omega}{\mathrm{d}k}=\pm \frac{c^2k}{\sqrt{c^2k^2+\omega_0^2}}=\pm \frac{c^2k}{\omega}= \pm \frac{c^2}{v_{\mathrm{ph}}}=\pm c\sqrt{1-\left(\frac{\omega_0}{\omega}\right)^2} \label{eq:vgr}
\end{equation}
For $\omega>\omega_0$, $k=\alpha$ and $\beta=0$, while for $\omega<\omega_0$, $k$ is imaginary ($k=i\beta$), meaning it denotes the evanescent wave, {\it i.e.} the wave varies sinusoidally in time, but exponentially in space. In the latter case, for a single component wave, $\alpha=0$, $v_{\mathrm{ph}}=\infty$, and $v_{\mathrm{gr}}=0$. With $t_{\mathrm{ph}}=\Delta z/v_{\mathrm{ph}}$ and $t_{\mathrm{gr}}=\Delta z/v_{\mathrm{gr}}$ it follows that for the evanescent wave we expect $t_{\mathrm{ph}}=0$ and $t_{\mathrm{gr}}=\infty$. However, as it is impossible to observe an infinite group travel time, we set the theoretical group travel times for the evanescent wave to zero. This is justified by the fact that in the case of the evanescent wave the whole solar atmosphere oscillates in phase, {\it i.e.} there is no detectable phase lag between different layers in the atmosphere. As we will see later, this approach agrees well with the observed dispersion relation in non-magnetic regions.  Note that under the assumption of an infinitely wide region where a wave is evanescent, the wave does not transmit energy vertically ({\it e.g.} Leibacher and Stein, 1981). However, if this regime is finite, tunneling can occur, in which case the wave does carry energy and there is phase propagation vertically.

The acoustic cut-off frequency $\omega_{\mathrm{0}}=\gamma g/4 \pi c$ is calculated from the latest solar atmosphere structures for the quiet Sun (Fontenla, Balasubramaniam, and Harder, 2007) and plage (Fontenla {\it et~al.}, 2006). Furthermore, $\gamma$=5/3 is the ratio of the specific heats, $g$=274\,m\,s$^{-2}$ the gravitational acceleration, $c =\sqrt{\gamma p/\rho }$ the sound speed, $p$ and $\rho$ the depth dependent pressure and density. Note that this is only one of several possible representations of the acoustic cut-off frequency in the quiet Sun (Schmitz and Fleck, 1998).
\subsection{Magnetic Waves ($a \gg c$)}
Let us now discuss the dispersion relation of upward traveling wave modes that can occur after mode transmission or conversion, which are the field guided slow acoustic mode and the fast magnetic mode. In the present study we focus on the question of whether we can identify the slow acoustic mode. This mode is understood to be ``leaking'' through magnetic regions by the lowering of the acoustic cut-off frequency in magnetic regions (see Bel and Leroy, 1977; Campos, 1987; De Pontieu, Erd{\'e}lyi and James, 2004; Cally, 2006). Here we account for the lowering of the acoustic cut-off frequency following Campos (1987, Equation 209a) and De Pontieu, Erd{\'e}lyi and James (2004):
\begin{equation}
\omega^2=c^2k^2+\omega_{\rm{eff}}^2, \label{eq:omega_eff}
\end{equation}
with $\omega_{\rm{eff}}=\omega_0\cos\theta$, and $\theta$ being the field inclination derived from the magnetic-field extrapolation. Note that this dispersion relation can again be derived from the dispersion relation given by Schunker and Cally (2006), Equation\,(12) for the limit of the magnetic case, {\it i.e.} $a \gg c$. Following the above outline for the acoustic wave, we get the phase and group velocity $v_{\mathrm{ph,sa}}$ and $v_{\mathrm{gr,sa}}$ for the field guided slow acoustic (sa) wave mode as:
\begin{equation}
v_{\mathrm{ph,sa}}=\frac{c}{\sqrt{1-\left(\frac{\omega_{\rm{eff}}}{\omega}\right)^2}} \label{eq:vph_acoustic}
\end{equation}
\begin{equation}
v_{\mathrm{gr,sa}}=c\sqrt{1-\left(\frac{\omega_{\rm{eff}}}{\omega}\right)^2}. \label{eq:vgr_acoustic}
\end{equation}

\section{Results}
In the following we compare the travel times derived from the observations with the theoretical travel times as outlined above. For consistency with the observations, we apply a running frequency filter to the theoretical travel times. This filter is the same Gaussian filter as used for the frequency filtering of the MOTH data ($\delta \omega=0.57$\,mHz). Figure\,\ref{fig:disp}b,c compares the observed travel times with the group and phase travel times, calculated as $t_{\rm ph}(\omega)=\Delta z/v_{\rm ph}(\omega)$, and $t_{\rm gr}(\omega)=\Delta z/v_{\rm gr}(\omega)$ for different regimes in the solar atmosphere. Panel (b) compares the observed travel times for $a < c$ with the theoretical values calculated with $\Delta z$=370\,km, $c=$6.5 km\,s${-1}$, $\omega_{\mathrm{0}}=5.6$\,mHz (see Table\,1) and no field inclination ($\theta=0^{\circ}$). The theoretical curve represents the observed dispersion relation quite well. For the intermediate-$\beta$ regime ($a\approx c$, panel c) we calculate the dispersion relation using $\Delta z$=220\,km, $c=$7.2\,km, and $\omega_{\mathrm{0}}=5.0$\,mHz for plage, and a lowered cut-off frequency with $\theta=30^{\circ}$ (dashed lines). We find that the fitted $t_{\mathrm{gr}}$ shows a substantially different behavior than the theoretical $t_{\mathrm{gr}}$. In particular, a cut-off frequency is not visible in the travel time fits from the observation. Therefore, we conclude that, while passing through the magnetic canopy, the acoustic nature of the wave is no longer detectable. 

\section{Discussion}
Let us discuss the significant change of the dispersion relation seen in Figure\,\ref{fig:disp}, panels (b) and (c). As the observation heights of Na and K are very close to the conversion layer, one might argue that we do not detect a pure slow acoustic wave with a lowered cut-off frequency, but a mixture of wave types, {\it i.e.} the slow acoustic and the fast magnetic. However, the clear correlation signal between both observing layers does not support this point. Therefore, it is striking that the acoustic dispersion relations seen in the high-$\beta$ regime vanishes {\it{immediately}} when the observing heights are close to the conversion layer. It is important to note here that the cross correlation yields robust results, so the wave packets lead to a clearly correlated signal in both observing layers. 

Furthermore, as the gas and magnetic pressure are of approximately equal amplitude, {\it i.e.} the magnetic field is weak, it is possible that the acoustic wave coming from below not only causes longitudinal pressure and velocity fluctuations along the field lines, but also distorts the magnetic-field lines along the wave vector (see Rosenthal {\it et~al.}, 2002). The combination of these processes might lead to a loss of the acoustic signature in the observations. 

One might expect to see the pure slow acoustic wave if we could discriminate waves propagating upward under an angle parallel to the magnetic field. In this case the acoustic wave can not distort the magnetic-field lines. Holography, as carried out by Schunker {\it et ~al.} (2008) for vertically upward traveling waves would be the right tool to select the appropriate incoming waves. Using this technique we might be able to analyze the dispersion relation of an 'ideal' slow acoustic wave.

The seemingly slight decrease of the travel times for the magnetic case (Figure\,\ref{fig:disp} dashed lines) for frequencies higher than 5\,mHz also needs further investigation. Muglach, Hofmann, and Staude (2005) conclude from MDI and TRACE observations that if the conversion layer is below the formation height of the TRACE intensity, the intensity power in the layers above decreases, which they associate with wave reflection. Numerical calculations carried out by Rosenthal {\it et~al.} (2002) show that wave reflection of high frequency waves (in their case $\omega=42$\,mHz) occurs at the $\beta=1$ layer. In the regions above the reflecting surface the wave is still present as an evanescent tail. This could explain the vanishing travel times  in the high frequency regime. However, only detailed 3D MHD simulations including realistic density and temperature stratifications can help to fully understand these phenomena.

\section{Conclusion}
From MOTH observation we show that in magnetic regions low-frequency waves can escape the solar acoustic cavity and propagate upwards between the K\,{\sc{i}} 7699\,{\AA} and Na\,{\sc{i}} 5890\,{\AA} observing layers. Furthermore, evidence is given that the observed low-frequency waves originate in deep layers as acoustic ($c\gg a$) and clearly change their characteristics while passing through the conversion layer into the low-$\beta$ regime. In particular, we do not see any evidence for a lowered cut-off frequency in low-$\beta$ regions. The reason for this is currently not explained and needs further investigation. 

It is however clear that in order to distinguish various wave types in the solar atmosphere, for a detailed analysis the observational heights have to be separated into the very high and low-$\beta$-regimes. This is essential for new instrument designs. Only by simultaneously cutting through very high and low-$\beta$ regimes can we analyze the characteristics of the waves in detail. As the different wave types dissipate their energy in different heights of the solar atmosphere, it is essential to determine to what extent the waves convert and transmit into the slow magnetic and fast acoustic wave. 

\begin{acks}
The authors very warmly thank C.~Lindsey for helpful discussions and comments. Furthermore, we thank S.~McIntosh for providing us with the code to perform the magnetic-field extrapolations, and J.D. Armstrong for providing an update of the Garbor wavelet fitting routine. We would also like to thank S.~Jefferies for detailed discussions and the hospitality during MH's visit at the IfA Hawaii, Maui. MDI data courtesy of the SOHO/MDI Consortium. SOHO is a project of international cooperation between ESA and NASA. The MOTH experiment was funded by award OPP-0087541 from the National Science Foundation (NSF). During this work MH was supported by the Swiss National Science Foundation under grant 200020-109420.
\end{acks}

%%% BIBLIOGRAPHY %%%%%%%%%%%%%%%%%%%%%%%%%%%%%%%%%%%%%%%%%%%%%%%%%%%%%%%%%%%
\bibliography{C:/Margit_Samsung/Texnic/bib1/ref_all03}

\appendix
In the following, we show how the dispersion relations given in Equations.\,(\ref{eq:omega}) and (\ref{eq:omega_eff}) can be derived from Equation~(12) in Schunker and Cally (2006). Let us first consider the acoustic wave, {\it i.e.} $a \rightarrow 0$. Under the assumption of vertically upward traveling waves, {\it i.e.} $k_x$=0, their equation simplifies to:
\begin{equation}
\omega^4-(a^2+c^2)k^2\omega^2+a^2c^2k^4\cos^2\theta-(\omega^2-a^2k^2\cos^2\theta)\omega_0^2=0. \label{eq:dispma}
\end{equation} 
Using the asymptotic case of $a=0$, Equation\,(\ref{eq:dispma}) further simplifies to

\begin{equation}
\omega^4-c^2k^2\omega^2-\omega^2\omega_0^2=0.
\end{equation}
After dividing with $\omega^2$, this then leads easily to Equation\,(\ref{eq:omega}). 
Now let us consider the asymptotic magnetic case with $a>>c$ or $1/a^2 \rightarrow 0$. Multiplying Equation\,(\ref{eq:dispma}) with $1/a^2$ we get

\begin{equation}
-k^2\omega^2+c^2k^4 \cos^2\theta +k^2 \cos^2\theta \omega_0^2=0.
\end{equation}
Dividing by $k^2$ and using $\cos^2\theta \omega_0^2=\omega_{\rm{eff}}^2$, and assuming vertically propagating waves leads to Equation\,(\ref{eq:omega_eff}).
\end{article} 

\end{document}